\documentclass[11pt]{article}

\usepackage{amsmath,amssymb}
\usepackage{epsfig}
\usepackage{graphicx}               
\usepackage{url}
\usepackage{hyperref}
\usepackage{bm}
\usepackage{cite}

\usepackage{color}

\newcommand{\nc}{\newcommand}

\nc{\beq}{\begin{equation}}
\nc{\eeq}{\end{equation}}
\nc{\beqa}{\begin{eqnarray}}
\nc{\eeqa}{\end{eqnarray}}
\nc{\bea}{\begin{eqnarray}}
\nc{\eea}{\end{eqnarray}}
\nc{\ra}{\rightarrow}
\nc{\lsim}{\begin{array}{c}\,\sim\vspace{-21pt}\\< \end{array}}
\nc{\gsim}{\begin{array}{c}\sim\vspace{-21pt}\\> \end{array}}
\nc{\Tr}{{\rm Tr}}
\nc{\slsh}{\slash\hspace*{-0.22cm}}

\def\be{\begin{equation}}
\def\ee{\end{equation}}
\def\bea{\begin{eqnarray}}
\def\eea{\end{eqnarray}}
\def\bit{\begin{itemize}}
\def\eit{\end{itemize}}

\def\slash#1{#1\!\!\!/\!\,\,}

\def\to{\rightarrow}

\title{
\vspace*{-2.3cm}
\begin{flushright}
\normalsize{
  }
\end{flushright}
\vspace{1.5cm}
\Large
\textbf{
A 750 GeV Dark Pion: Cousin of a Dark G-parity-odd WIMP
}\vspace*{1.0cm}
}

\author{ {\bf Yang Bai}, {\bf Joshua Berger} and {\bf Ran Lu}
\vspace{5mm}
\\
\normalsize\emph{Department of Physics, University of Wisconsin-Madison, Madison, WI 53706, USA}  
}

\date{}

\begin{document}
\setcounter{page}{0}
\maketitle

\vspace*{1cm}
\begin{abstract}
We point out a potential common origin of the recently observed 750 GeV diphoton resonance and a Weakly Interacting Massive Particle (WIMP) candidate. In a dark QCD sector with an unbroken dark $G$-parity, the diphoton resonance could be a dark $G$-even pion, while the WIMP could be the lightest dark $G$-odd pion. Both particles are Standard Model gauge singlets and have the same decay constant. For the dark pion decay constant of around 500 GeV, both the diphoton excess at the LHC and the dark matter thermal abundance can be accommodated in our model. Our model predicts additional dark $G$-even and dark $G$-odd color-octet pions within reach of the 13 TeV LHC runs. For the $5+\bar{5}$ model, compatible with the Grand Unified Theories, the WIMP mass is predicted to be within $(613, 750)$~GeV. 
\end{abstract}

\thispagestyle{empty}
\newpage

\setcounter{page}{1}

\baselineskip18pt

\vspace{-3cm}

\section{Introduction}
\label{sec:intro}
Strongly coupled composite sectors are a generic prediction of
asymptotically free gauge theories.  As such, they have long played a
role in particle physics models, not the least of which is determining
the phenomenology of hadrons.  In the context of physics beyond the
Standard Model (BSM), they can also provide interesting and unique
signals at the Large Hadron Collider (LHC) \cite{Kilic:2009mi} and can
alleviate the hierarchy problem~\cite{Hill:2002ap}.  The bound states that arise in
strongly coupled sectors can have accidental discrete symmetries that
suppress or eliminate their decays.  For example, if the constituent
``quarks'' are vector-like, transform under representations of a $SU(2)$
flavor symmetry and otherwise fall into real representations of any
non-Abelian gauge groups, they can be protected by a new $G$-parity
\cite{Bai:2010qg}, which is a generalization of charge conjugation
symmetry.  An approximate version of such a symmetry protects light
hadrons from decaying in QCD, but is broken by electroweak interactions.  A new
strongly coupled composite sector could also exhibit its own
$G$-parity.  If the symmetry is exact, then the lightest $G$-odd
particle is stable and is a candidate for dark matter. We will call
the sector with new strong dynamics similar to QCD the dark QCD sector.

Recent results \cite{CMS-PAS-EXO-15-004,ATLAS-CONF-2015-081} from Run 2 at the LHC have hinted at a new diphoton resonance with a mass of 750 GeV.  If the data are really displaying the first signs of BSM physics at the LHC, then they point to a relatively large $\sigma \times {\rm Br} \sim 5{\rm -}10~{\rm fb}$ for the new resonance into diphoton.  The large couplings of the neutral resonance to gluons and photons hint at a new strongly coupled sector, of which the 750 GeV particle could be a pion.  Such a scenario has been considered independently  in Refs.~\cite{Low:2015qep,Molinaro:2015cwg,Ellis:2015oso,DiChiara:2015vdm,Franceschini:2015kwy,Pilaftsis:2015ycr,Buttazzo:2015txu,Nakai:2015ptz,Harigaya:2015ezk}, but in this work we extend such models to a case where the new strongly coupled sector (dark QCD) has an exact dark $G$-parity and demonstrate that this model has a viable dark matter candidate as the lightest $G$-odd pion.  Other potential explanations of the LHC diphoton excess have been presented in Refs.~\cite{Gupta:2015zzs,Bellazzini:2015nxw,McDermott:2015sck,Higaki:2015jag,Knapen:2015dap,Angelescu:2015uiz,Backovic:2015fnp,Mambrini:2015wyu}.

The dark $G$-odd pion, as a cousin of the 750 GeV diphoton resonance, can annihilate into photons and gluons, so that it can be a thermal relic dark matter. Its couplings to photons and gluons are related to the 750 GeV diphoton resonance couplings, because both particles arrive from the same strongly coupled sector. We demonstrate below that these interactions could lead to the stable dark pion making up either a significant fraction or all of the observed dark matter.  The same interactions generate potential signals for dark matter direct and indirect detection, as well as mono-X type signals at the LHC.

In addition, the relative couplings of the $G$-even dark pion, the 750 GeV diphoton resonance, to other gauge bosons are predicted within the model with no free parameters.  In particular, we provide concrete predictions for $Z\gamma$, $ZZ$, $WW$ and $gg$ observations of this unstable dark pion.  The remaining dark pions in this model couple directly to QCD or electroweak gauge bosons and could lead to spectacular dijet or multi-particle signals.

The rest of this paper is structured as follows.  In Section~\ref{sec:models}, we present the $5+\bar{5}$ dark QCD model and determine its basic properties in Section~\ref{sec:model-55}, compatibility with the LHC diphoton excesses in Section~\ref{sec:diphoton-55}, and dark matter potential in Section~\ref{sec:WIMP}. We briefly discuss the other $8+1+1$ model in Section~\ref{sec:model-8+1+1} and conclude in Section~\ref{sec:conclusion}.

\section{Dark QCD Models with a Dark $G$-parity}
\label{sec:models}
As noted in Ref.~\cite{Bai:2010qg}, one could have an unbroken $G$ parity in the dark QCD sector under the assumption that the dark quarks have real representations under the SM gauge interactions. For irreducible representations under the SM gauge group, no $G$-even dark pions can explain the 750~GeV diphoton resonance, at least for models in Ref.~\cite{Bai:2010qg}. Beyond the simplest models, one could choose dark quark representations to have both a $G$-odd dark pion for dark matter and a $G$-even dark pion for the diphoton resonance. In this section, we present one representative model to discuss the correlation between the dark matter phenomenology and the diphoton-related signals. 

\subsection{The $5 + \bar{5}$ Model}
\label{sec:model-55}
To be compatible with the Grand Unified Theories, it is natural to consider the $SU(5)_{\rm GUT}$ representations. Furthermore, to conserve the $G$-parity, we need to have a ``real" representation of $SU(5)_{\rm GUT}$, so we introduce ${5} + {\bar{5}}$ as the first example.~\footnote{We also note the possibility of combining ${5} + {\bar{5}}$ into a real representation, 10, of $SO(10)$.  A common Dirac mass for $\psi_1$ and $\psi_2$ can be explained. } The particle content in terms of dark quarks is listed in Table~\ref{tab:fieldcontent-55}. 
\begin{table}[ht!]
\renewcommand{\arraystretch}{1.8}
\begin{center}
\begin{tabular}{ccc}
\hline \hline
   &   $SU(N_d)_{\rm dQCD}$    &   $SU(5)_{\rm GUT}$    \\  \hline
$\psi_{1\, L,R} $    &  $N_d$          &  5                      \\ \hline
$\psi_{2\, L,R}$    &  $N_d$          &  $\overline{5}$                       \\ \hline  \hline
\end{tabular}
\end{center}
\caption{Field content of a model with a confining dark QCD gauge group $SU(N_d)_{\rm dQCD}$.
\label{tab:fieldcontent-55}}
\end{table}%

The basic Lagrangian is 
\beqa
{\cal L} &=& {\cal L}_{\rm SM} 
-\frac14 (\hat{F}^a_{\mu\nu})^2 
+ \bar{\psi}_1 \left( i\,\slash{\partial} \,+ \,\hat{g}\,\slash{\hat{A}}^b\,\hat{t}^{b} \,+ \,g_5\,\slash{A}^a\,t^{a}  \right)\psi_1
+ \bar{\psi}_2 \left( i\,\slash{\partial} \,+ \,\hat{g}\,\slash{\hat{A}}^b\,\hat{t}^{b} \,- \,g_5\,\slash{A}^a\,t^{a*}  \right)\psi_2
  \nonumber \\
&& \hspace{1cm} \, -\, ( \bar{\psi}_1\,{\cal M}_\psi \, \psi_1 +  \bar{\psi}_2\,{\cal M}_\psi \, \psi_2) \,. 
\label{eq:Lagrangian-5+5}
\eeqa
Here, the dark QCD gauge field is denoted by $\hat{A}_\mu$ with the
generator as $\hat{t}^b$; the GUT $SU(5)$ gauge field is denoted by
$A_\mu^a$ with the generator as $t^a$ with $\mbox{Tr}[t^a
t^b]=\frac{1}{2}\delta^{ab}$. We also introduce a mass matrix, ${\cal
  M}_\psi$, for both $\psi_1$  and $\psi_2$. At the scale around 1
TeV, in order to conserve the SM gauge symmetry, there are two
parameters for this mass matrix
\beqa
{\cal M}_\psi = \mbox{diag}(m_1 \, \mathbb{I}_3, m_2 \, \mathbb{I}_2)\,,
\eeqa
with $m_1, m_2 \geq 0$ in our convention. 

The above Lagrangian is invariant under the ``dark $G$-parity'', under which the dark quarks, dark gauge fields and the GUT gauge fields transform as
\beqa
\psi_1 &\xrightarrow{G}& \psi^{\cal C}_2 = i\,\gamma^2\,\psi^*_2 \,, \qquad \qquad \quad 
\psi_2 \xrightarrow{G} \psi^{\cal C}_1 = i\,\gamma^2\,\psi^*_1 \,, 
 \nonumber \\
\hat{A}^b\,\hat{t}^b &\xrightarrow{G}& (\hat{A}^b)^{\cal C}\,\hat{t}^b = \hat{A}^b\,(-\hat{t}^{b*}) \,, 
\qquad  A_\mu^a \xrightarrow{G} A_\mu^a  \,.
\label{eq:G-55}
\eeqa
with $\cal C$ denoting charge conjugation.  Note in particular that $A^a_\mu$, as well as all remaining SM fields are invariant. It is readily verified that the 
Lagrangian in Eq.~(\ref{eq:Lagrangian-5+5}) is invariant under 
the $G$-parity transformation in Eq.~(\ref{eq:G-55}).  Hence dark $G$-parity is a good quantum number of the theory, and all SM particles are $G$-even.

In order to keep asymptotic freedom for both $SU(N_d)_{\rm dQCD}$ and SM QCD gauge couplings, we further require $2 \leq N_d \leq 5$. The $SU(N_d)_{\rm dQCD}$ gauge coupling becomes strong in the infrared and the confinement and chiral symmetry breaking happen at a scale $\Lambda_d \approx 4\pi f_\Pi$, with $f_\Pi$ as the dark pion decay constant. For $m_1, m_2 \lesssim \Lambda_d$ and in the low energy theory below $\Lambda_d$, we have totally 99 PNGB's, dark pions. Decomposing them into $SU(5)$ representations, we have
\beqa
{\bf 10} \times {\bf 10} - {\bf 1} &=& \bm{24}^+ + \bm{24}^- + \bm{1}^- + \bm{10} + \bm{15} + \overline{\bm{10}} + \overline{\bm{15}}  \,.\label{eq:10-decompose}
\eeqa
Here, the superscript ``$+(-)$" means the $G$-parity even(odd).~\footnote{For the model based on 10 of $SO(10)$, the decomposition of 99 dark pions under $SO(10)$ is ${\bf 10} \times {\bf 10} - {\bf 1} = \bm{45}^- + \bm{54}^+$.} We can further decompose these pions into SM gauge group representations. For instance, using the notation of $\left(SU(2)_W, SU(3)_{\rm QCD}\right)_{U(1)_Y}$ one has
\beqa
\bm{24}^+ &=& (\bm{1}, \bm{1})_{\bm{0}}^+ + (\bm{3}, \bm{1})_{\bm{0}}^+ + (\bm{2}, \bm{3})_{\bm{-5/6}}^+ + (\bm{2}, \bm{\bar{3} })_{\bm{5/6}}^+ + (\bm{1}, \bm{8} )_{\bm{0}}^+ \,, \nonumber \\
\bm{24}^- &=& (\bm{1}, \bm{1})_{\bm{0}}^- + (\bm{3}, \bm{1})_{\bm{0}}^- + (\bm{2}, \bm{3})_{\bm{-5/6}}^- + (\bm{2}, \bm{\bar{3} })_{\bm{5/6}}^- + (\bm{1}, \bm{8} )_{\bm{0}}^- \,.
\label{eq:su5-decompose}
\eeqa
There are totally three SM gauge singlets with one $G$-parity even and two $G$-parity odd. We will denote $(\bm{1}, \bm{1})_{\bm{0}}^+$ simply as $\bm{1}^+$ and $(\bm{1}, \bm{1})_{\bm{0}}^-$ simply as $\bm{1}^-_A$, while the singlet in Eq.~(\ref{eq:10-decompose}) as $\bm{1}^-_B$. Their normalized generators are
\beqa
T^{\bm{1}^+} &=& \frac{1}{2\,\sqrt{30}} \, \mbox{diag}( 2\,\mathbb{I}_3, - 3\,\mathbb{I}_2, 2\,\mathbb{I}_3, - 3\,\mathbb{I}_2 ) \,, \nonumber \\
T^{\bm{1}^-_A} &=& \frac{1}{2\,\sqrt{30}} \, \mbox{diag}( 2\,\mathbb{I}_3, - 3\,\mathbb{I}_2, - 2\,\mathbb{I}_3,  3\,\mathbb{I}_2 ) \,, \nonumber \\
T^{\bm{1}^-_B} &=& \frac{1}{\sqrt{20}} \, \mbox{diag}( \mathbb{I}_5, -\mathbb{I}_5 ) \,,
\eeqa
which has the canonical normalization with $\mbox{Tr}(T^{\bm{1}^+} T^{\bm{1}^+} ) =1/2$ and the same for other pion generators. To calculate couplings of dark pions with SM gauge bosons, we will use the $10\times 10$ matrix representation of the $SU(5)$ generators as $T^a = \mbox{diag}( t^a, - t^{a*})$ with $\mbox{Tr}(T^a T^a) = \delta^{ab}$. To complete the decomposition into SM gauge group representation, we also have $\bm{10} = (\bm{1}, \bm{1})_{\bm{1}} + (\bm{1}, \bm{\bar{3}})_{\bm{-2/3}} + (\bm{2}, \bm{3})_{\bm{1/6}}$ and $\bm{15} = (\bm{3}, \bm{1})_{\bm{1}} + (\bm{2}, \bm{3})_{\bm{1/6}} + (\bm{1}, \bm{6})_{\bm{-2/3}}$.

\begin{figure}[th!]
\begin{center}
\includegraphics[width=0.6\textwidth]{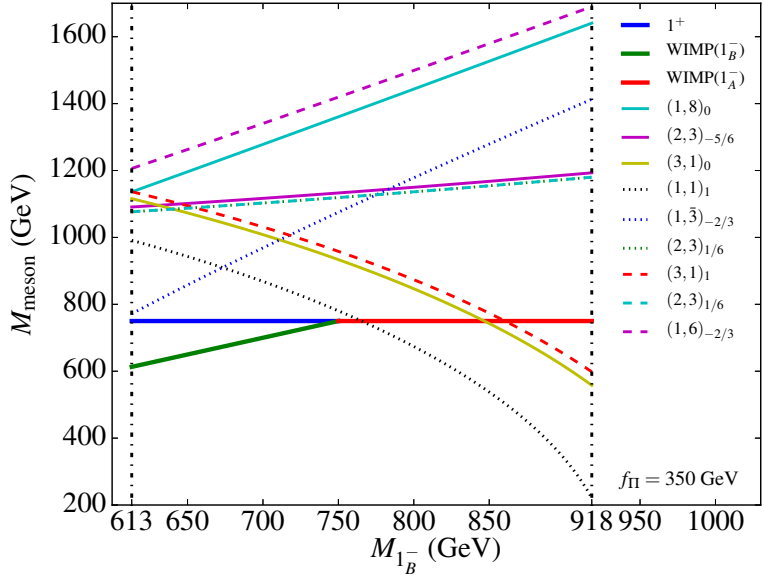}
\caption{Dark pion masses as a function of the $\bm{1}^-_B$ mass after fixing the lightest $G$-even dark pion, $\bm{1}^+$, to be 750 GeV and $f_\Pi = 350$~GeV. The QCD gauge couplings is chosen to have $\alpha_s(1~\mbox{TeV})\approx 0.09$. We don't show the masses of additional dark pions related to the ones in this plot by complex conjugation.}
\label{fig:mass-spectra}
\end{center}
\end{figure}

Since the SM gauge interactions can provide masses for charged pions, we anticipate that the three SM-singlet pions are lighter than others. The bare dark quark masses contribute to the dark pions via the general formula
\beqa
M^2_{AB} = \frac{1}{f^2_\Pi}\, \langle \overline{Q}\, \{ T^A, \{ T^B, M_Q \} \} \, Q \rangle_0 \,.
\label{eq:pion-mass-general}
\eeqa
with $A,B=1,\cdots,10$ and $M_Q = \mbox{diag}({\cal M}_\psi, {\cal M}_\psi)$. The dark quark condensation can be used to define the dark QCD chiral symmetry breaking scale via $\langle \overline{Q} Q \rangle =  \Lambda_d^3/(16\pi^2)\,\mathbb{I}_{10}$. Applying the general formula in Eq.~(\ref{eq:pion-mass-general}) to our model, we have 
\beqa
M^2_{\bm{1}^+} = M^2_{\bm{1}^-_A}  = \left( \frac{4}{5} m_1 + \frac{6}{5} m_2 \right) \, \frac{\langle \overline{Q} Q \rangle}{f^2_\Pi}  \,,  \qquad 
M^2_{\bm{1}^-_B}  = \left( \frac{6}{5} m_1 + \frac{4}{5} m_2 \right) \, \frac{\langle \overline{Q} Q \rangle}{f^2_\Pi}   \,.
\eeqa
The contributions to other SM charged dark pion masses are proportional to $2 m_1$, $2 m_2$, or $m_1 + m_2$, depending on their charges. For SM charged dark pions, the contributions from SM gauge boson loops are quadratically divergent and are
\beqa
\Delta M^2[(d_2, d_3)_Y] \approx \frac{3}{16\pi^2} \Lambda_d^2 
\left[ g_s^2 C_2(d_3) + g_2^2 C_2(d_2) + g_Y^2 Y^2 \right] \,.
\eeqa

For a benchmark point of $f_\Pi = 350$~GeV and fixing $M_{\bm{1}^+}=750$~GeV, we show various dark pion masses as a function of $M_{\bm{1}^-_B}$ in Fig.~\ref{fig:mass-spectra}. For $0 \leq m_1 < m_2$, the lightest $G$-odd particle is ${\bm{1}^-_B}$, while for $0 \leq m_2 < m_1$ the lightest $G$-odd particles is ${\bm{1}^-_A}$. As a result and after fixing $M_{\bm{1}^+}=750$~GeV, our model predicts a range for the potential WIMP mass as
\beqa
{\bf 613~\mbox{\bf GeV} \leq M_{\rm \bf WIMP} \leq 750~\mbox{\bf GeV} }\,. 
\eeqa
We also note that there is an upper mass $M_{\bm{1}^-_B} \leq 918$~GeV, saturated by choosing $m_2=0$.

\subsection{Properties of ${\bf 1}^+$: 750 GeV diphoton resonance}
\label{sec:diphoton-55}
For the $G$-even dark pion, the triangular anomalies mediate its interactions with two gluons and two photons. The general form for the anomaly-mediated interactions are
\beqa
{\cal L}_{\rm anomaly} = - \frac{g_B g_C\, \Pi_A}{16\pi^2\, f_\Pi} \epsilon_{\mu\nu\rho\sigma} F^{\mu\nu}_B F^{\rho\sigma}_C \,\mbox{Tr}\{ T^A T^B T^C \} \,+\,\cdots \,.
\eeqa
Applying this general form to $\bm{1}^+$ in our model, we have
\beqa
{\cal L}_{\rm anomaly}& = & - \frac{N_d}{\sqrt{30} }\, \frac{g_s^2 }{16\pi^2\, f_\Pi} \,\Pi_{\bm{1}^+}\,\epsilon_{\mu\nu\rho\sigma} G^{\mu\nu\,a} G^{\rho\sigma\,a}
+ \frac{3\,N_d}{2\,\sqrt{30} }\, \frac{g_2^2 }{16\pi^2\, f_\Pi} \,\Pi_{\bm{1}^+}\,\epsilon_{\mu\nu\rho\sigma} W^{\mu\nu\,i} W^{\rho\sigma\,i} \nonumber \\
&&+ \frac{N_d}{2\,\sqrt{30} }\,\frac{\left(\sqrt{\frac{5}{3}}g_Y\right)^2}{16\pi^2\, f_\Pi} \,\Pi_{\bm{1}^+}\,\epsilon_{\mu\nu\rho\sigma} B^{\mu\nu} B^{\rho\sigma}
 \,. 
\eeqa
For this model, we have several decay channels for $\bm{1}^+$. We show the branching fractions and the total widths in Table~\ref{table:branching-55}. 
\begin{table}[ht!]
 \renewcommand{\arraystretch}{2.0}
  \centering
\begin{tabular}{|c|c|c|c|c|c|}
\hline\hline
Mode &  $gg$  & $\gamma\gamma$ &  $Z \gamma$ & $Z Z$ & $W W$ \cr \hline
Branching ratio & 0.90 & 0.0046 & 0.0083 & 0.021 &  0.065 \cr \hline
$\Gamma_{\rm tot}$ & \multicolumn{5}{c|}{ $0.13~\mbox{GeV}\,\left( \frac{N_d}{4} \right)^2 \left( \frac{500~\mbox{GeV} }{f_\Pi} \right)^2$  } \cr
\hline\hline
\end{tabular}
\caption{The decay branching ratios of $\bm{1}^+$ and its total width for the $5+\bar{5}$ model with a mass of 750 GeV.}
\label{table:branching-55}
\end{table}
We note that neglecting the phase space factors, the ratio of the branchings of $\gamma\gamma$, $Z \gamma$ and $ZZ$ is
\beqa
\mbox{Br}(\gamma\gamma) : \mbox{Br}(Z \gamma) : \mbox{Br}(Z Z) = 1 : \frac{ (9 \cos^2{\theta_W} - 5 \sin^2{\theta_W})^2 }{98 \,\cos^2{\theta_W}\,\sin^2{\theta_W} } :  \frac{(9 \cos^4{\theta_W} + 5 \sin^4{\theta_W})^2 }{196 \,\cos^4{\theta_W}\,\sin^4{\theta_W} }  \,,
\label{eq:branching-55}
\eeqa
with $\theta_W$ as the weak mixing angle and $\sin^2{\theta_W} \approx 0.23$. 

\begin{figure}[th!]
\begin{center}
\includegraphics[width=0.65\textwidth]{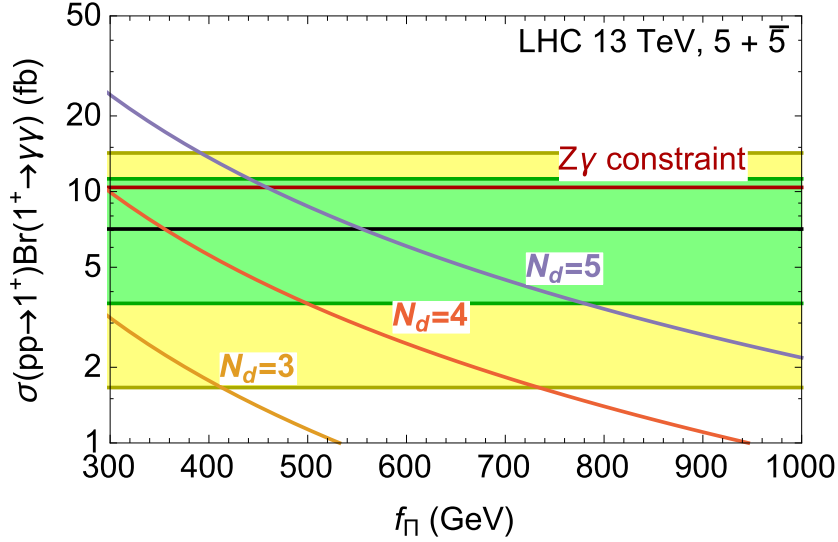}
\caption{The production cross section  times branching ratio for $pp \rightarrow \bm{1}^+ \rightarrow \gamma\gamma$ at the 13 TeV LHC, with $M_{\bm{1}^+}=750$~GeV. The $\sigma\times \mbox{Br}$ for the 750 GeV diphoton excess is shown in the black line with $1\sigma$ (green) and $2\sigma$ (yellow) bands. We also show the rescaled $Z\gamma$ constraint from the 8 TeV results~\cite{Aad:2014fha}. }
\label{fig:diphoton-production-55}
\end{center}
\end{figure}

Using the MSTW 2008 PDFs~\cite{Martin:2009iq} and the narrow-width approximation, we calculate the production cross section times diphoton branching ratio in Fig.~\ref{fig:diphoton-production-55} as a function of $f_\Pi$. For the diphoton excess, we estimate a range for the allowed values of the cross-section by combining the CMS and ATLAS 13 TeV results.  We consider the number of observed and expected background predicted in the respective analysis papers by summing over the two bins nearest to 750 GeV.  Using the signal acceptances determined by simulations in the experimental papers, we determine that the maximum likelihood cross section for the combined CMS and ATLAS excesses is 7.1 fb, with $1\sigma$ ranges of $[3.6, 11.2]$ fb and $2\sigma$ ranges of $[1.7, 14.2]$ fb.  The maximum likelihood cross-section is indicated by a black line and the 1$\sigma$ and 2$\sigma$ lines by green and yellow bands respectively in Fig.~\ref{fig:diphoton-production-55}. One can see that for $N_d$ from 3 to 5 and $f_\Pi$ from 300~GeV to 1~TeV, the 750 GeV diphoton excess can be explained in our model.  

A prior search using data from Run 1 constrained the production of a
$Z \gamma$ resonance with a mass of 750 GeV.  They obtained a limit at this mass of
\begin{equation}
  \label{eq:1}
  \sigma(pp \to \bm{1}^+) \, {\rm Br}\left[\bm{1}^+ \to(Z \to \ell\ell)  \gamma \right] <
  0.27~{\rm fb}\,,\qquad (8~{\rm TeV}) \,,
\end{equation}
from Ref.~\cite{Aad:2014fha}.
Using the well-known leptonic branching fraction of the $Z$ to
leptons, ${\rm Br}(Z \to \ell^+\ell^-) = 0.0673$, the $gg$ parton
luminosity ratio at 750 GeV from 13 TeV to 8 TeV of 4.7, and the
branching fraction ratios determined from Eq.~(\ref{eq:branching-55}), we can relate this to a constraint on 13 TeV production of $\gamma\gamma$ via the
$\bm{1}^+$ resonance.  We find
\begin{equation}
  \label{eq:2}
  \sigma(pp \to \bm{1}^+)\, {\rm Br}(\bm{1}^+ \to \gamma\gamma) < 9.6~{\rm
    fb}\,,\qquad (13~{\rm TeV})\,.
\end{equation}
We show this constraint line in
Fig.~\ref{fig:diphoton-production-55}. The majority of the parameter
space that offers an explanation of the diphoton excess remains allowed
after imposing this constraint.

\subsection{Properties of a ${\bf 1}^-$ WIMP}
\label{sec:WIMP}
As discussed above, the lightest $G$-odd particle, $\bm{1}_A^-$ or
$\bm{1}_B^-$, could serve as a candidate for WIMP dark matter. If $\bm{1}_B^-$ is lighter than $\bm{1}_A^-$, the WIMP mass could be from 613 GeV to 750 GeV. If $\bm{1}_A^-$ is lighter, the WIMP mass is fixed to be 750 GeV, the same as the 750 GeV diphoton resonance. Here we don't consider the multi-component dark matter case with a degenerate mass for $\bm{1}_A^-$ and $\bm{1}_B^-$. For both cases, the WIMP is a SM singlet and does not have renormalizable interactions with SM particles. On the other hand, because of the composite nature of ${\bf 1}^-_{A, B}$, it has higher-dimensional operator interactions with SM gauge fields. Specifically, we have three relevant operators for both cases
\beqa
&& c_G \,\frac{g_s^2\,N_d}{16\pi^2 f_\Pi^2}\, \Pi_{\bm{1}^-}\Pi_{\bm{1}^-} \, G^{\mu\nu\,a} G_{\mu\nu}^a \,+\, c_W \,\frac{g_2^2\,N_d}{16\pi^2 f_\Pi^2}\, \Pi_{\bm{1}^-}\Pi_{\bm{1}^-} \, W^{\mu\nu\,i} W_{\mu\nu}^i  \nonumber \\
&& \,+\, c_B \,\frac{\left(\sqrt{\frac{5}{3}}g_Y\right)^2\,N_d}{16\pi^2 f_\Pi^2}\, \Pi_{\bm{1}^-}\Pi_{\bm{1}^-} \, B^{\mu\nu} B_{\mu\nu} \,.
\label{eq:DD-operator}
\eeqa
The first operator is the dark matter chromo-Rayleigh interaction~\cite{Godbole:2015gma,Bai:2015swa} and the second and third operators are related to the electromagnetic polarizability of dark matter ~\cite{Weiner:2012cb,Appelquist:2015zfa}. Up to an order-of-unity number, $\eta$, from non-perturbative physics, we have
\beqa
&&\hspace{-1cm}\bm{1}^- = \bm{1}_B^-: \quad  613~\mbox{GeV} \leq  M_{\bm{1}^-} < 750~\mbox{GeV},  \qquad c_G = c_W = c_B = \frac{1}{20} \eta\,, \\
&&\hspace{-1cm}\bm{1}^- = \bm{1}_A^-: \quad   M_{\bm{1}^-} = 750~\mbox{GeV}, \qquad c_G = \frac{1}{30}\eta \,, c_W = \frac{3}{40}\eta \,, c_B = \frac{7}{120}\eta \,.
\eeqa

The WIMP candidate, ${\bf 1}^-$, mainly annihilates to two gluons with the annihilation rate 
\beqa
\langle \sigma v \rangle ({\bf 1}^- {\bf 1}^- \rightarrow G G ) = \frac{4\,c_G^2\,N_d^2\,\alpha_s^2}{\pi^3} \, \frac{M^2_{{\bf 1}^-} }{f_\Pi^4} + {\cal O}(v^2) \, \approx \, 1.0~\mbox{pb}\cdot\mbox{c} \,,
\label{eq:thermal}
\eeqa
for $c_G =\eta/20$ with $\eta = 1.3$, $N_d =4$, $M_{{\bf 1}^-} = 650$~GeV and $f_\Pi=350$~GeV. This has the right amount of annihilation rate to provide a thermal WIMP.

\begin{figure}[th!]
\begin{center}
\includegraphics[width=0.65\textwidth]{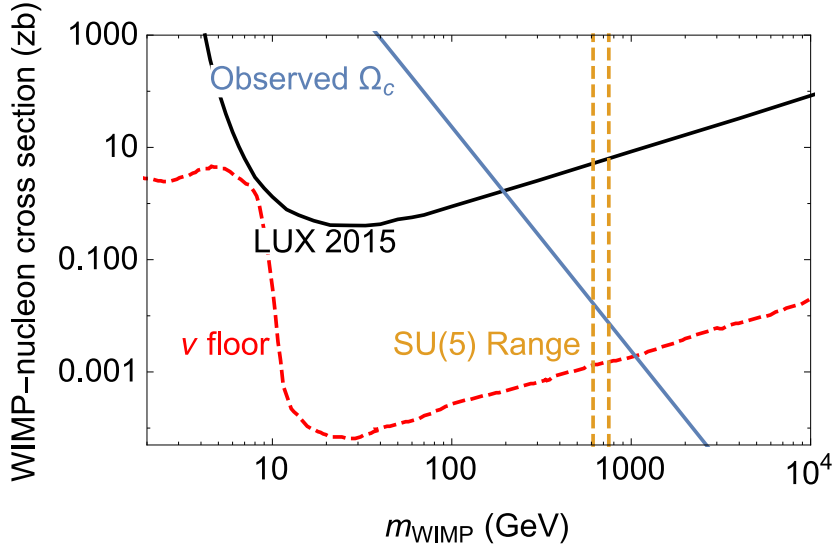}
\caption{The blue line is our model predictions of the spin-independent scattering cross sections after satisfying the dark matter thermal relic abundance $\Omega_c$. For the $5 + \bar{5}$ model, the WIMP mass is predicted to be within $(613, 750)$~GeV. The dashed and red line is the ``neutrino floor" cross section~\cite{Billard:2013qya}.
}
\label{fig:direct-detection}
\end{center}
\end{figure}

For direct detection experiments and following a standard calculation like in Ref.~\cite{Bai:2015swa}, we have the spin-independent scattering cross section as 
\beqa
\sigma^{\rm SI}_{{\bf 1}^- N}  \,= \, \frac{\kappa^2\,c_G^2\,N_d^2\, m_N^4}{4\pi f_\Pi^4 (m_N + M_{{\bf 1}^-})^2 } \, =\,1.4 \times 10^{-47}~\mbox{cm}^2  \,,
\label{eq:direct}
\eeqa
for $c_G =\eta/20$ with $\eta = 1.3$, $N_d =4$, $M_{{\bf 1}^-} = 650$~GeV and $f_\Pi=350$~GeV. Here, $m_N$ is the nucleon mass and $\kappa \approx -0.20$ and is related to the matrix element of gluon operators inside a nucleon. We note that the predicted scattering cross section in our model is below the current bound from LUX~\cite{Akerib:2015rjg} and could be probed by future direct detection experiments. Furthermore, taking the ratio of Eq.~(\ref{eq:thermal}) and Eq.~(\ref{eq:direct}), one can see that only the dark matter mass $M_{{\bf 1}^-}$ becomes the relevant parameter for the direct detection predictions in our model. In Fig.~\ref{fig:direct-detection}, we show our model predictions for WIMP spin-independent scattering cross sections together with the LUX limits. The predicted cross sections for our $5+\bar{5}$ model are around two orders of magnitude below the current constraints.

\subsection{The $8+1+1$ Model}
\label{sec:model-8+1+1}
Before we finish this section, we briefly mention other possible
models with no obvious compatibility with a GUT theory. For instance,
we could have a ``$8+1+1$" model with one QCD octet dark quarks plus
two dark quarks with opposite hypercharges. Their charges under the SM
gauge group is listed in Table~\ref{tab:fieldcontent}. We note that
for $N_d \geq 2$, SM QCD is not asymptotic free above the dark
confinement scale and has a Landau-pole scale depending on $N_d$.  The
resulting SM QCD Landau pole scale can be above the Planck scale only
for $N_d = 2$. 

\begin{table}[ht!]
\renewcommand{\arraystretch}{1.8}
\begin{center}
\begin{tabular}{ccccc}
\hline \hline
   &   $SU(N_d)_{\rm dQCD}$    &   $SU(3)_c$    &   $SU(2)_W$    & $U(1)_Y$    \\  \hline
$\psi_{L,R}$    &  $N_d$          &  8                 &    1                  & $0$      \\ \hline
$\chi_{1\, L,R}$    &  $N_d$          &  1                 &    1                  & $x$      \\ \hline  
$\chi_{2\, L,R}$    &  $N_d$          &  1                 &    1                  & $-x$      \\ \hline  \hline
\end{tabular}
\end{center}
\caption{Field content of the $8+1+1$ model.  
\label{tab:fieldcontent}}
\end{table}%
The basic Lagrangian is 
\beqa
{\cal L} &=& {\cal L}_{\rm SM} 
-\frac14 (\hat{F}^a_{\mu\nu})^2 
+ \bar{\psi}\left( i\,\slash{\partial} \,+ \,\hat{g}\,\slash{\hat{A}}^b\,\hat{t}^{b} \,+ \,g_s\,\slash{G}^a\,t^{a}  \right)\psi \nonumber \\
&& 
\,+\, \bar{\chi}_1\left( i\,\slash{\partial} \,+ \,\hat{g}\,\slash{\hat{A}}^b\,\hat{t}^{b} \,+\, x\,g_Y\,\slash{B} \right)\chi_1 \,+ \,
 \bar{\chi}_2\left( i\,\slash{\partial} \,+ \,\hat{g}\,\slash{\hat{A}}^b\,\hat{t}^{b}  \,-\, x\,g_Y\,\slash{B}\right) \chi_2 \nonumber \\
&& \,-\, m_1\, \bar{\psi} \psi  \, -\, m_2\, ( \bar{\chi}_1 \chi_1 +  \bar{\chi}_2 \chi_2) \,. 
\label{eq:Lagrangian-8+1+1}
\eeqa
Here, the SM QCD gauge field is denoted by $G_\mu^a$ with the generator as $t^a$ with $\mbox{Tr}[t^a t^b]=C(8)\,\delta^{ab}=3\,\delta^{ab}$. The dark $G$-parity is defined as
\beqa
\psi &\xrightarrow{G}& \psi^{\cal C} = i\,\gamma^2\,\psi^* \,, \qquad 
\chi_1 \xrightarrow{G} \chi_2^{\cal C} = i\,\gamma^2\,\chi_2^* \,, \qquad
\chi_2 \xrightarrow{G} \chi_1^{\cal C} = i\,\gamma^2\,\chi_1^*  \nonumber \\
\hat{A}^b\,\hat{t}^b &\xrightarrow{G}& (\hat{A}^b)^{\cal C}\,\hat{t}^b = \hat{A}^b\,(-\hat{t}^{b*}) \,, 
\qquad  G^a_\mu \xrightarrow{G} G^a_\mu  \,,
\qquad  B_\mu \xrightarrow{G} B_\mu  \,.
\label{eq:G}
\eeqa

In the low energy theory below $\Lambda_d$, we have totally 99 PNGB's, which are
\beqa
{\bf 10} \times {\bf 10} - {\bf 1} &=& \bm{8}_{\rm A}^- + \bm{10}^- + \overline{\bm{10}}^- + \bm{8}_{\rm S}^+ + \bm{27}^+ + \bm{8}_{x} +  \bm{8}_{-x} + \bm{8}_{x} +  \bm{8}_{-x}  \nonumber \\
&&\hspace{-3mm} + \, \bm{1}_{2x} + \bm{1}_{-2x} + \bm{1}^+ + \bm{1}^- \,.
\eeqa
Here, the subscript denotes the electric charge of dark pions; ``A" and ``S" mean anti-symmetric and the symmetric combinations. Applying the general formula in Eq.~(\ref{eq:pion-mass-general}) to our model, we have 
\beqa
M^2_{\bm{1}^+}  = \left( \frac{2}{5} m_1 + \frac{8}{5} m_2 \right) \, \frac{\langle \overline{Q} Q \rangle}{f^2_\Pi}  \,,  \qquad 
M^2_{\bm{1}^-}  = 2\,m_2 \, \frac{\langle \overline{Q} Q \rangle}{f^2_\Pi}   \,.
\eeqa
%
In this model, $\bm{1}^-$ is the lightest $G$-odd dark pion in the spectrum. In the latter study, we just treat $M_{\bm{1}^+} $ and $M_{\bm{1}^-} $ as two free parameters.

\begin{figure}[th!]
\begin{center}
\includegraphics[width=0.65\textwidth]{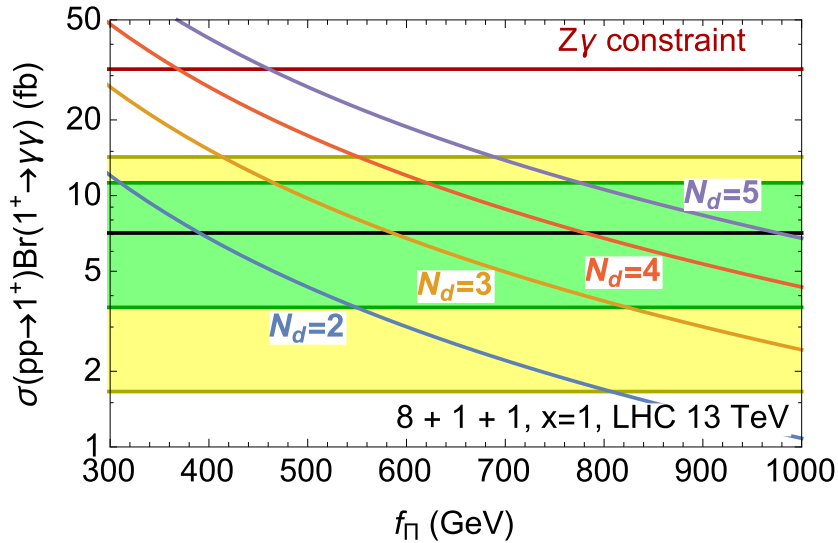}
\caption{The same as Fig.~\ref{fig:diphoton-production-55} but for the $8+1+1$ model with $x=1$. }
\label{fig:diphoton-production}
\end{center}
\end{figure}

The ${\bf 1}^+$ dark pion can serve as the 750 GeV diphoton resonance. Its anomalous couplings to gauge bosons are
\beqa
{\cal L}_{\rm anomaly} = - \frac{3\,N_d}{4\sqrt{5} }\, \frac{g_s^2 }{16\pi^2\, f_\Pi} \,\Pi_{\bm{1}^+}\,\epsilon_{\mu\nu\rho\sigma} G^{\mu\nu\,a} G^{\rho\sigma\,a}
+ \frac{x^2\,N_d}{\sqrt{5} }\, \frac{g_Y^2 }{16\pi^2\, f_\Pi} \,\Pi_{\bm{1}^+}\,\epsilon_{\mu\nu\rho\sigma} B^{\mu\nu} B^{\rho\sigma}
 \,. 
\eeqa
It has the ratios of the branchings of $\gamma\gamma$, $Z \gamma$ and $ZZ$ as
\beqa
\mbox{Br}(\gamma\gamma) : \mbox{Br}(Z \gamma) : \mbox{Br}(Z Z) = 1 : \frac{2 \sin^2{\theta_W} }{\cos^2{\theta_W} } :  \frac{\sin^4{\theta_W} }{\cos^4{\theta_W} } \,.
\label{eq:branching}
\eeqa
In Fig.~\ref{fig:diphoton-production}, we show the production cross section times diphoton branching ratio as a function of $f_\Pi$ for the benchmark model point $x=1$. One can see that for $N_d$ from 2 to 5 and $f_\Pi$ from 400~GeV to 1~TeV, the 750 GeV diphoton excess can be explained in this model. The derived constraint from the 8 TeV $Z\gamma$ resonance searches is less than 32 fb and much weaker than that for the $5+\bar{5}$ model.  

The $G$-odd dark pion ${\bf 1}^-$ is a stable particle and could be a
WIMP candidate. The calculation is very similar to the $5+\bar{5}$
model except with different coefficients in
Eq.~(\ref{eq:DD-operator}). For this $8+1+1$ model, one has $c_G = \frac{3}{2}
\eta$, $c_W =0$ and $c_B = \frac{x^2}{2} \eta$. After satisfying the
WIMP thermal relic abundance, the predicted direct detection cross
sections for different WIMP masses are also as shown in
Fig.~\ref{fig:direct-detection}. The LUX experiments have already
imposed a bound on the dark matter mass to be above $\sim 160$~GeV.

\section{Discussion and Conclusions}
\label{sec:conclusion}
The model described in this work has additional pions that could be
observed at the LHC. For the $5+ \bar{5}$, many QCD-charged pions have a mass around 1 TeV. Pair productions of them at the LHC will lead to interesting signatures with or without missing energy. For instance, the $G$-even and color-octet,  $(\bm{1}, \bm{8})^+_{\bm{0}}$ or $\bm{8}_S^+$, appears as high mass pair-produced dijet resonances at $m > 750~{\rm GeV}$. For the $8+1+1$ model, the $\bf{10}$ and $\bf{27}$ receive large color factor enhancements and give spectacular signals. In addition, there are hypercharged octets, which may require
large multiplicity final states in order to decay depending on their
hypercharges.  We defer a detailed discussion of their
phenomenology.  

Since the WIMP candidate, $\bm{1}^-$, can annihilate into a pair of gluons, which will eventually hadronize, and produce secondary particles including photons, positrons, antiprotons and neutrinos. Those particles can give dark matter indirect detection signals. On the other hand, if we require the WIMP thermal relic abundance match to the observed dark matter energy density, the annihilation cross section is too small to be constrained by the current dark matter indirect searches. Besides the dominate gluon annihilation channel, our WIMP candidate can also annihilate into a pair of photons, generating a monochromatic gamma ray signal. This signal is suppressed by $\alpha_s^2/\alpha_Y^2 \sim {\cal O}(100)$, so the current experiments are not sensitive to it. It is still an interesting signal for future indirect dark matter search experiments.

In summary, we have constructed a dark QCD model with an unbroken dark $G$-parity. The 750 GeV diphoton resonance could be explained by the dark $G$-even and SM gauge singlet pion. In the same model, the lightest dark $G$-odd pion provides a good WIMP candidate.

\subsection*{Acknowledgments}
This work is supported by the U. S. Department of Energy under the contract DE-FG-02-95ER40896.

\bibliography{750}
\bibliographystyle{JHEP}
 \end{document}